\documentclass[12pts,preprint]{aastex}
\usepackage{emulateapj5}
\usepackage{onecolfloat5}
\usepackage{natbib}
\bibpunct{(}{)}{;}{a}{,}{,}

\input epsf



\long\def\comment#1{}

\def\W2{{\cal W}}

\def\be{\begin{equation}}
\def\ee{\end{equation}}
\def\bea{\begin{eqnarray}}
\def\eea{\end{eqnarray}}

\def\Mpc{\,{\rm Mpc}}

\def\cmm2{{\,\rm cm^{-2}}}
\def\cm2{{\,{\rm cm}^2}}
\def\cmm3{{\,{\rm cm}^{-3}}}
\def\gcmm3{{\,{\rm g\,cm^{-3}}}}

\def\fun#1#2{\lower3.6pt\vbox{\baselineskip0pt\lineskip.9pt
  \ialign{$\mathsurround=0pt#1\hfil##\hfil$\crcr#2\crcr\sim\crcr}}}

\def\C{{\cal C}}
\begin{document}
\bibliographystyle{apj}
\twocolumn[
\title{Normal parameters for an analytic description of the CMB cosmological
parameter likelihood}
\author{Mike Chu, Manoj Kaplinghat and Lloyd Knox}
\affil{Department of Physics, University of California,
Davis, CA 95616, USA, email:
chu@physics.ucdavis.edu,kaplinghat@ucdavis.edu,lknox@ucdavis.edu}

\begin{abstract}
 The normal parameters are a non--linear transformation of the
 cosmological parameters whose likelihood function is very
 well--approximated by a normal distribution. This transformation
 serves as an extreme form of data compression allowing for
 practically instantaneous calculation of the likelihood of any given
 model, as long as the model is in the parameter space originally
 considered.  The compression makes all the information about
 cosmological parameter constraints from a given set of experiments
 available in a useable manner.  Here we explicitly define the normal
 parameters that work for the current CMB data, and give their mean
 and covariance matrix which best fit the likelihood function
 calculated by the Monte Carlo Markov Chain method.  Along with
 standard parameter estimation results, we propose that future CMB
 parameter analyses define normal parameters and quote their mean and
 covariance matrix.
\end{abstract}
 ]


\section{Introduction}

The challenge of turning a CMB dataset into cosmological parameter
constraints is one that has been solved by a series of data
compressions.  The information in the time--ordered data is
compressed into a map \citep{wright96,tegmark97a}.  The information
in the map is then compressed into a power spectrum
\citep{tegmark97b,bond98,bond00,wandelt01,bartlett00,hivon02}. Finally,
the information in the power spectrum is then compressed into
cosmological parameters \citep{lineweaver97,benoit02}.

This last step, however, is more of a data ``explosion'' than a
data compression. Although the number of parameters is indeed
small ($\sim 10$) the non--Gaussian distribution of their errors
may be characterized by $10^{10}$ numbers for grid--based
likelihood evaluation \citep{tegmark00} or perhaps as little as
$30,000$ for the Monte-Carlo Markov Chain method
\citep{christensen01}.

This explosion has undesirable consequences.  The full
constraining power of the data is not used by the cosmological
community.  Typically authors report not the full (cumbersome)
likelihood, but projections or marginalizations of it down to one
or at most two--dimensional spaces. Given the near--degeneracies
that exist in the probability distribution, and its non-Gaussian
nature, these final steps lose significant information.

Our work here provides the final step of compression in the data
analysis pipeline.  We compress the probability distribution of
the cosmological parameters down to the $\sim 50$ parameters of
an analytic form for the likelihood (a mean and covariance matrix).
This step makes the full information in the parameter likelihood function
easily useable.

Our procedure is analogous to the `radical compression'\citep{bond00}
used for compressing the uncertainty in the power spectrum
estimates, whose distribtuion is also non--Gaussian.  In
both cases a non--linear variable transformation is used, with the
property that the transformed variables are well--approximated by
a normal distribution.

We were inspired to search for a set of normal parameters by a
proposal that the $C_l$ can have a
linear dependence on a set of well-chosen parameters that span the space of
possible models \citep{kosowsky02}.  Clearly such a parameter
set would be tremendously valuable for parameter estimation since the $C_l$
for any given model could be calculated practically instantaneously simply
by summing the terms in the first order Taylor expansion. Such a set
would also, given sufficiently Gaussian errors on $C_l$,  have errors
with a normal distribution.

Although inspired by \citet{kosowsky02}, our goal here is entirely
different.  It is to find a set of parameters (the normal parameters)
for which the likelihood is Gaussian.  Although the approximate linearity
demonstrated by \citet{kosowsky02} gave us confidence to try to find a
set of normal parameters, it by no means guaranteed success. The
linear-response approximation has not yet been demonstrated to be
sufficiently accurate for purposes of parameter estimation
given MAP data (and certainly not given current data) and
the errors on $C_l$ are not actually Gaussian.

While \citet{kosowsky02} are developing a tool for parameter
estimation from CMB data, our work is aimed at being able to
easily take full advantage of such parameter estimation.  The most
straightforward application of CMB data compressed in this form is
for simultaneous analysis of CMB constraints with those from other
cosmological probes.  In the past to do this analysis one would
have to entirely reproduce the reduction from bandpowers (plus
their window functions, offset log--normal parameters, Fisher
matrices, and calibration uncertainties) to cosmological parameter
likelihoods. This procedure requires the organization of a lot of
data as well as hundreds of thousands of angular power spectrum
calculations.  In contrast, with the compression to normal
parameters in hand one need only perform a variable transformation
and evaluate a $~10$--dimensional Gaussian.  We use a similar set
of parameters as KMJ, and demonstrate explicitly the validity of
the normal approximation by numerical computation of the
likelihood given current CMB data.

One can also simplify a probability distribution by choosing linear
combinations of the parameters that diagonalize the covariance matrix,
the so--called parameter eigenmodes \citep{efstathiou99}.  However, the
resulting parameters are still dependent and non--Gaussion if the
distribution of the original parameters is non-Gaussian. We do find
diagonalization of the covariance matrix {\em of the normal parameters} to
be useful, both for our fitting procedure and for gaining an understanding
of the nature of the constraints the data places on the parameter space.

We test our procedure on current data and provide the mean and
covariance matrix to the normal parameters that best fit the
likelihood given current data.  As we go to press, this fit is now
out of date due to the data from the Wilkinson Microwave
Anisotropy Probe ({\it WMAP}) \footnote{http://map.gsfc.nasa.gov}.
However, we expect these data to further improve the validity of
the normal approximation to the likelihood and thus our method
will be of lasting value.

In section II we describe the normal parameters. In section III we
describe our calculation of the exact likelihood via the Monte
Carlo Markov Chain method \citep{christensen01}.  In section IV we
describe our procedure for finding the mean and covariance matrix
that provide the best fit to our MCMC--calculated likelihood. In
section V we present our results. In section VI we discuss
applications and conclude.

\section{The normal parameters}

The normal parameters are a non-linear combination of the
six cosmological parameters that we consider here -- $\omega_b$,
$\omega_d$, $\Omega_{\Lambda}$, $z$, $A\equiv (2.73\times 10^6)^2 \Mpc^{-3}
P_\Phi(0.05\Mpc^{-1})$ and
$n_s$ (baryon density, dark matter density, dark energy density,
reionization redshift, amplitude and scalar tilt of primordial power
spectrum respectively).  In particular, we replace $\Omega_{\Lambda}$,
$A$ and $n_s$ by
\begin{eqnarray}
\Theta_s&\simeq & \Theta_s^E \equiv r_s/d_A \\
A^*& \equiv &{A\over 76,000} \,\left({0.05\Mpc^{-1}\over
k_{\rm pivot}}\right)^{1-n_s}e^{-2\tau}  \\
t&\equiv&\frac{1}{\sqrt{w_b}}\, 2^{n_s-1}
\end{eqnarray}
where $r_s$ is the sound horizon at recombination, $d_A$ is the
angular diameter distance to the recombination surface and we take
$k_{\rm pivot}=0.067 \Mpc^{-1}$. The distinction between
$\Theta_s$ and $\Theta_s^E$ is explained below.  The numerical
factors are chosen so that $A^*$ is of order unity.  We use $\tau$
to denote the optical depth to Thomson scattering from here back
to some time before reionization and after recombination.  The
resulting variable set, $\{\omega_b, \omega_d, \Theta_s, A^*, t,
z\}$, has a probability distribution, given CMB data, that is
well--approximated by a Gaussian as we will demonstrate below.

We chose the normal parameters by considering what combinations of
cosmological parameters affect the features in the angular power
spectrum over the range in which the data are highly constraining
($100 \la l \la 500$).  The parameter $A^*$ describes the overall
amplitude, $t$ is chosen to correlate with the ratio of the amplitude
of the second peak to the first peak (at fixed $\omega_m$), and
$\Theta_s$ scales the features horizontally.

\citet{hu01} have investigated this phenomenology as well.  For
$\Theta_S$ we use their approximation to the angular size of the sound
horizon, which has the advantage of being an algebraic
expression\footnote{We set $\Theta_s \equiv \pi/l_A$ where $l_A$ is
given by their equations A3-A5.}, rather than the exact sound horizon,
$\Theta_s^E$.  This distinction is important because they can differ
by more than the uncertainty in $\Theta_S$.  \citet{hu01} also define
a parameter $H_2$ analogous to our $t$ but with a different parameter
dependence.

The importance of $\Theta_s$ for understanding the $C_l$ is widely
recognized
\citep{efstathiou99,tegmark00,hu01,kaplinghat02,kosowsky02}.  It
is one of the best--determined cosmological quantities:
$\Theta_s = 0.^\circ 59 \pm 0.^\circ 01$ \citep{knox01}.

The parameter $A^*$ sets the overall amplitude of the spectrum at
$l\ga 100$.  At these $\ell$ values, Thomson scattering depresses the
amplitude by $\exp(-2\tau)$.  The value of $k_{\rm pivot}$ was chosen
to decorrelate $A^*$ and $n_S$.  In an idealized case
with uniform relative error bars on $l(l+1)C_l/(2\pi)$ from some
$l_{\rm min}$ to $l_{\rm max}$ we would expect $k_{\rm pivot}=\sqrt{l_{\rm min} l_{\rm max}}/\eta_0$ where $\eta_0 = (14 \pm 0.6)$ Gpc is approximately
the coordinate distance to the last--scattering surface. Taking $l_{\rm min}=10$
and $l_{\rm max}=1000$ we expect $k_{\rm pivot} \sim 0.007 \Mpc^{-1}$.
At fixed $\tau$ (or $z$) we find this value works well.  Allowing $\tau$ to
vary leads to a correlation with $n_S$.  A result of this correlation
is that decorrelation between $A^*$ and $n_S$ is best done with a much
larger $k_{\rm pivot}$ of about $0.067 \Mpc^{-1}$.
The $\tau$-$n_S$ correlation arises because both rely considerably
on $C_l$ measurements at $l < 100$.

The least ``normal'' of the normal parameters is $z$, which is not
very well--constrained by CMB data.  The situation may improve quite
soon with large--angle polarization data from MAP
\citep{kaplinghat02a}.  Or, it is already not a problem if one
interprets observation of a Gunn--Peterson trough in a $z=6.3$ quasar
as indicating $z_{\rm RI} \simeq 6.3$ \citep{becker01,fan02}.  We
caution that though the data indicate the amount of neutral Hydrogen
in the inter--galactic medium is rising from zero, how much the
fraction of free electrons is increasing with $z$ is highly
unconstrained \citep{kaplinghat02a}.

We assume a step function transition of the ionization fraction
from 0 to 1 at redshift $z$.  
For the flat models we consider this results in an optical depth of 
\be
\tau = 0.038\omega_b h /\omega_m \left[ \sqrt{\Omega_\Lambda + \Omega_m(1+z)^3}-1 \right]
\ee
\citep{hu97,kaplinghat02a}.  We settled on $z$ as a normal
parameter after trial and error.  We have found $z$ to be more
normal than either $\tau$ or $\exp{-2\tau}$.

We have not included tensor perturbations, curvature or dark
energy models with $w \equiv p/\rho \ne -1$ in our analysis and so
our results strictly only apply with these assumptions.  Including
all these variations (at fixed $\Theta_s$) would only alter the
$C_l$ at $l \la 60$.  Their inclusion would therefore affect our
constraints on $z$ and $n_S$ (and therefore $t$), but not $A^*$,
$\Theta_s$, $\omega_b$ or $\omega_d$.

\section{Likelihood Calculation}

What we want to know is, given the data and any other assumptions we make
about the world, what is the probability distribution of the parameters?
This posterior probability distribution can be calculated by use of
Bayes' theorem which states:
\be
P({\vec \theta}|d) \propto P(d|{\vec \theta}) P_{\rm prior}({\vec \theta})\,.
\ee
where $d$ refers to data and the proportionality constant is chosen to
ensure  $\int P({\vec \theta}|d) d {\vec \theta}=1$.
With a uniform prior this simply reduces to
$P({\vec \theta}|d) \propto P(d|{\vec \theta})$.
This probability of the data given the parameters is, when thought of
as a function of the parameters, called the likelihood,
${\cal L}({\vec \theta})$.

Often we are interested in the posterior probability distribution
for one or two parameters alone.  This marginalized posterior
is given by integrating over the other parameters.  For example:
\be
P(\theta_1,\theta_2|d)= \int \Pi_{i=3}^n d\theta_i {\cal L}(\vec \theta)
P_{\rm prior}(\vec \theta)
\ee
where $n$ is the number of parameters.  We use the
prior to incorporate non--CMB information such as that the redshift of
reionization must be greater than 6.3 \citep{becker01}.

In the following  subsections we discuss first how we evaluate the
likelihood function at a single point and then how we evaluate it over
a large parameter space and produce marginalized posterior
distributions.

\subsection{Likelihood evaluation}

Here we take the data to be the measured averages of the angular power
spectrum (called bandpowers), $D_i^d$.  The expected signal contribution
to $D_i^d$ is given by an average over the power spectrum:
\be
\label{eqn:bandpower}
D^s_i = \sum_l u^2_{\alpha(i)}f_{il}\C_l(\vec\theta)
\ee
where $u_{\alpha(i)}$ is the calibration
parameter for dataset $\alpha$ and
$\C_l \equiv l(l+1)C_l/(2\pi)$ is the angular power spectrum.

The uncertainty in the $D_i$ is non--Gaussian but well--approximated by
the offset log--normal form of \citet{bond00}.
Specifically, ${\cal L} = \exp(-\chi^2/2)$ where
\begin{eqnarray}
\label{eqn:chisq}
\chi^2 & =& \sum_{i,j}\left(Z_i^{\rm s}-Z_i^{\rm d}\right)M_{ij}^Z
\left(Z_j^{\rm s}-Z_j^{\rm d}\right) + \chi^2_{\rm expt}; \\
\chi^2_{\rm expt} &\equiv & \sum_\alpha {(u_\alpha-1)^2
\over \sigma_\alpha^2}+ (b-\bar b)^2/\sigma_b^2; \\
Z_i^{\rm d} & \equiv & \ln(D_i^d+x_i); \\
Z_i^{\rm s} &\equiv&
\ln\left(D_i^s + x_i\right); \\
M_{ij}^Z & \equiv & M_{ij}\left(D_i^d+x_i\right)\left(D_j^d+x_j\right)\quad
\mbox{no sum};
\end{eqnarray}
where $M_{ij}$ is the weight matrix for the band power data $D_i^d$.
The expt label is for experimental parameters.  These include the
calibration parameters and a beam--width parameter, $b$, for the
one experiment with significant and quantified uncertainty in their beam width.
For simplicity, we take the prior probability
distribution for the experimental parameters to be normally distributed.  Since the
datasets have already been calibrated, the mean of the calibration parameters
is at $\bar u_{\alpha}=1$.  The calibration parameter index, $\alpha$, is a
function of $i$ since different power spectrum constraints from the
same dataset all share the same calibration uncertainty.

We include bandpower data from Boomerang, the Degree Angular Scale
Interferometer \citep[DASI;][]{halverson02}, Maxima \citep{lee01}),
the COsmic Background Explorer\citep[COBE;][]{bennett96}, VSA and CBI.
The weight matrices, band powers and window functions are publicly
available for both VSA and DASI.  For COBE we approximate the window
functions as tophat bands; all other information is available in
\citet{bond00} and in electronic form with the DASh package.  For
Boomerang, CBI and Maxima we approximate the window functions as
top-hat bands, the weight matrices as diagonal and the log-normal
offsets, $x$, as zero. The Boomerang team report the uncertainty in
their beam full-width at half-maximum (fwhm) as $12.9 \pm 1.4$ minutes
of arc. We follow them in modelling the departure from the nominal
(non-Gaussian) beam shape as a Gaussian. For the Boomerang $D_i^s$, we
replace $\C_l$ in Eq.~\ref{eqn:bandpower} with $\C_l \exp(-l^2
b^2)$. We do not allow the fwhm to go below 11'.5 or above 14'.3; i.e.
we give such fluctuations zero probability.  To reduce our sensitivity
to beam errors, for Boomerang and Maxima we only use bands with
maximum $l$-values less than 1000. For CBI we use their broader
mosaic'ed fields with bandpowers that extend to $l=1900$.

\subsection{Exploring the Parameter Space}

Our first step in exploring the high--dimensional parameter space
is the creation of an array of parameter values called a chain,
where each element of the array, $\vec \theta$, is a location in
the $n$-dimensional parameter space.  The chain has the useful
property once it has converged that $P(\vec \theta \in R) =
N(\vec \theta \in R)/N$  where the left--hand side is the
posterior probability that $\vec \theta$ is in the region $R$, $N$
is the total number of chain elements and $N(\vec \theta\in R)$ is
the number of chain elements with $\vec \theta$ in the region $R$.
Once the chain is generated one can then rapidly explore
one--dimensional or two--dimensional marginalizations in either
the original parameters, or in derived parameters, such as $t_0$.
Calculating the marginalized posterior distributions is simply a
matter of histogramming the chain.

The chain we generate is a Monte Carlo Markov Chain (MCMC) produced
via the Metropolis--Hastings algorithm described in Christensen et
al. (2001).  The candidate--generating function for an initial run was
a normal distribution for each parameter.  Subsequent runs
used a multivariate--normal distribution with cross--correlations
between cosmological parameters equal to those of the posterior as
calculated from the initial run.

The covariance matrix of the candidate--generating function, $C_G$ is actually
a scaled version of the posterior covariance matrix, $C_P$: $C_G= a C_P$.
Proper choice of $a$ is important if the movement through parameter
space is to be efficient.  If $a$ is too low, the acceptance rate will be
high, but the typical step size will be small and a full sampling of the
parameter space will take a long time.  If $a$ is too high then most steps
will be to regions of greatly reduced probability, the acceptance rate
will therefore be low and once again a full sampling of the parameter space
will take a long time.  As a useful measure of the speed at which we
sample the parameter space, we define the step distance to be:
\begin{equation}
\Delta \sigma_{n,n-1}  =\sqrt{ \sum_i
\Delta \theta_i  C_{P,ij}^{-1} \Delta \theta_j    };
\end{equation}
where $\Delta \theta_i$=$(\theta_n)_i-(\theta_{n-1})_i$, $n$ is the iteration
number and $i$ enumerates the parameters.  In other words we are using the
inverse covariance matrix as the distance metric for the parameter space.
We automatically adjust $a$ in the first 20,000 samples of a run to
maximize this average  step distance.  We find it has a fairly broad
plateau between 0.3 and 0.7.   In practice we calculate the square of
the distance with the slightly simpler expression:
$\sum_i (\Delta \theta_i)^2/C_{P,ii}$.

All of our results are based on an MCMC run consisting of 754,000
iterations. For the ``burn-in'' the initial 20,000
samples were discarded, and the remaining set was thinned by accepting
every 25th iteration, resulting in nearly 30,000 samples.
The acceptance rate was 43\% with $a=.37$ and an average step distance of 0.25
indicating highly efficient movement through the parameter space.
We used the CODA software \citep{best95} to confirm
that the chain passed the Referty-Lewis convergence
diagnostics and the Heidelberger-Welch stationarity test.

\section{Fitting to the Likelihood}

We do not fit directly to the ten--dimensional likelihood or even
to its marginalization down to the six cosmological parameters.
The reason is that in the high--dimensional spaces, the likelihood
is very sparsely sampled.  For a fairly coarse grid with 8 steps in each direction,
there would be $6^8 > 250,000$ grid points and yet we have only 30,000 samples.
Only with the marginalizations down to even lower--dimensions does the
likelihood become densely sampled.

Instead we fit to the two--dimensional marginalized likelihoods and from
these fits reconstruct the full result.  We typically use 50 bins in
each parameter which results in about 100 samples in the most-likely
of the 2500 bins.

We do this fitting in the normal parameter space.  That is, at each point
in the chain we calculate the value of the two normal parameters of interest
and histogram that point accordingly.  For every such pair of normal
parameters we find $\bar \theta_i$ and $F^{2d}_{ij}$ that give the best--fit
two--dimensional Gaussian likelihood where
\be
{\cal L}_{2d}^N \propto
\exp\left[-{1 \over 2} \left(\theta_i - \bar
  \theta_i\right)F^{2d}_{ij}\left(\theta_j -
\bar \theta_j\right)\right].
\ee
From these $n \times (n-1)/2$ ($= 15$ for $n=6$) fits we reconstruct
the six--dimensional Gaussian.  We exploit a property of Gaussian
distributions that the inverse of  $F^{2d}_{ij}$ gives the elements
of the covariance matrix in the full higher--dimensional space:
$C_{ij}=(F^{2d})^{-1}_{ij}$.  Thus we uniquely determine the 15 independent
off-diagonal elements of the 6 by 6 covariance matrix.  The diagonal
elements are over--determined with five estimates of $C_{ii}$ for each $i$; we
average these together for the final estimate.  The full covariance matrix is then
inverted, giving the 6--dimensional Fisher-matrix of the likelihood
function.  The $\bar \theta_i$ are also averaged to produce an
estimate of the likelihood maximum.

Note that each of the off-diagonal elements is obtained from a 2D
gaussian fit to different variable pairs. Therefore there is no
guarantee that the covariance matrix built from these 2x2 blocks will
necessarily be positive-definite (which a true covariance matrix
should be). We employ an iterative procedure to solve the above
problem. We first increase the diagonal terms of $C$ until $C$ is
positive-definite. We then invert $C$ to get the Fisher matrix
$F=C^{-1}$ and find its eigenmodes. Now we identify these eigenmodes
as the ``new normal parameters'' and perform the 2D fit procedure
again (this step is not computationally intensive). The resulting
covariance matrix (from the 2D fits) will be close to diagonal and
hence positive definite. We repeat this procedure typically about 15
times by which time the changes in the covariance matrix are less than
3\%. We have also verified that the final result is insensitive to the
arbitrary adjustments to the covariance matrix (to make it
positive-definite) in the first step.

The end result is an approximate likelihood, ${\cal L}^N$, with most likely values $\bar \theta_i$
and covariance matrix $C_{ij}$:
\be
-2 \ln{{\cal L}^N} = \left(\theta_i - \bar \theta_i\right)C^{-1}_{ij} \left(\theta_j - \bar \theta_j\right).\ee
If we denote the cosmological parameters with $\vec c$ and the normal parameters with $\vec \theta$, the
likelihood of the cosmological parameters is given by
\be
{\cal L}^c(\vec c)  =  {\cal L}^N(\vec \theta(\vec c))
|{\partial \vec{c} \over \partial \vec{\theta}}|
\ee
where
\be
|{\partial \vec{c} \over \partial \vec{\theta}}| =  {\partial
  \Theta_s \over \partial \Omega_\Lambda}{A^* \over A}\left({0.05 \Mpc^{-1}
  \over k_{\rm pivot}}\right)^{n_s -1} b\ln 2
\ee
is the Jacobian of the variable transformation.

\section{Results}

In Table~\ref{tab:normfit} we show the parameters of ${\cal L}^N$
that make it a good approximation to our MCMC--calculated likelihood.
For ease of interpreting the magnitude of the off--diagonal elements we
have shown the correlation matrix which is given by $C_{ij}/\sqrt{C_{ii} C_{jj}}$
instead of $C_{ij}$ itself.

\begin{figure}[!ht]
\centerline{\scalebox{.4}{\includegraphics{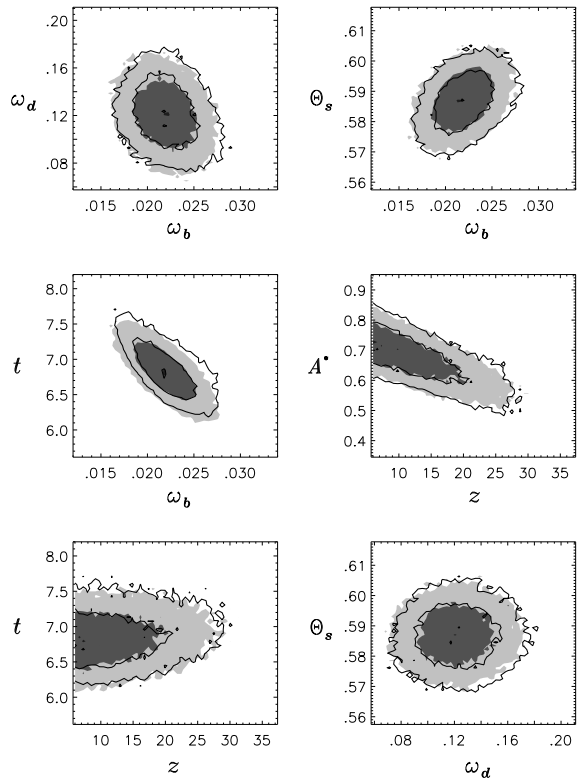}}}
\caption[]{\label{fig:2dnorm} Comparison of the two--dimensional
marginalized posterior probability distributions of the normal parameters
for the normal fit (shading) and the MCMC--calculated likelihood (contours).  Contours
and shading show where the probability is above $\exp(-2.3/2)$ and
$\exp(-6.17/2)$ of the maximum value.
}
\end{figure}

We compare the fit to the exact likelihood in Fig.~\ref{fig:2dnorm}
which shows 6 of the 15 possible marginalizations down to two
dimensions.  The marginalization of the normal fit is done by a Monte
Carlo process.  We use the normal fit to rapidly generate a chain
whose elements are samples from the normal distribution.  We can then
manipulate this chain (e.g., to marginalize down to 2 dimensions) just
like we manipulate the Markov chains.

\clearpage

\clearpage

\begin{figure}[!ht]
\centerline{\scalebox{.4}{\includegraphics{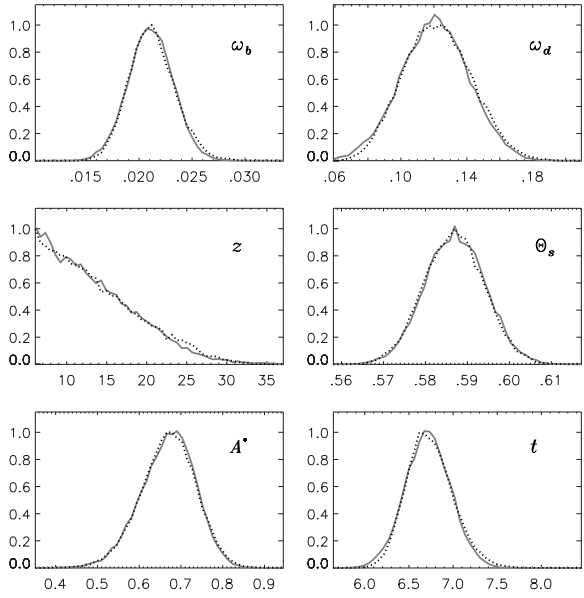}}}
\caption[]{\label{fig:1dnorm}Comparison of the one--dimensional
marginalized posterior probability distributions of the normal
parameters for the normal fit (solid, red) and the MCMC--calculated
likelihood (dotted, black).
}
\end{figure}

All six marginalizations down to one dimension are shown in
Fig.~\ref{fig:1dnorm}.  One can see that the fit provides an excellent
match to the exact one--dimensional likelihoods.

Note that although the fit is Gaussian in the six--dimensional space,
the priors implicit in the marginalization down to one dimension
result in slightly non--Gaussian one--dimensional likelihoods.  This
is because the marginalization is not done over the full parameter
space but has the constraints $z > 6$, $0 < \Omega_\Lambda < 0.9$,
$0.45 < h < 0.95$ and $0.02 < \Omega_m < 1.1$.

\clearpage

\clearpage

\begin{figure}[!ht]
\centerline{\scalebox{.4}{\includegraphics{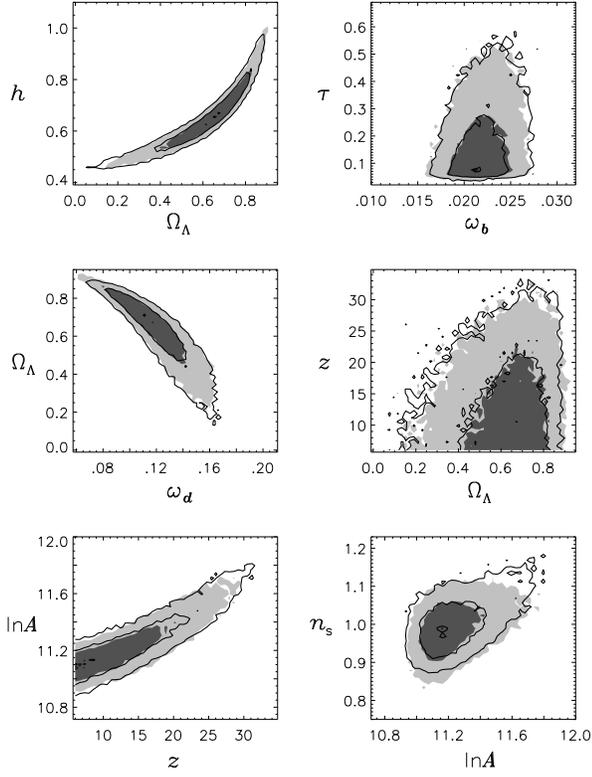}}}
\caption[]{\label{fig:2dcosmo}As in Fig.~\ref{fig:2dnorm} but
for other parameters.}
\end{figure}

\begin{figure}[!ht]
\centerline{\scalebox{.4}{\includegraphics{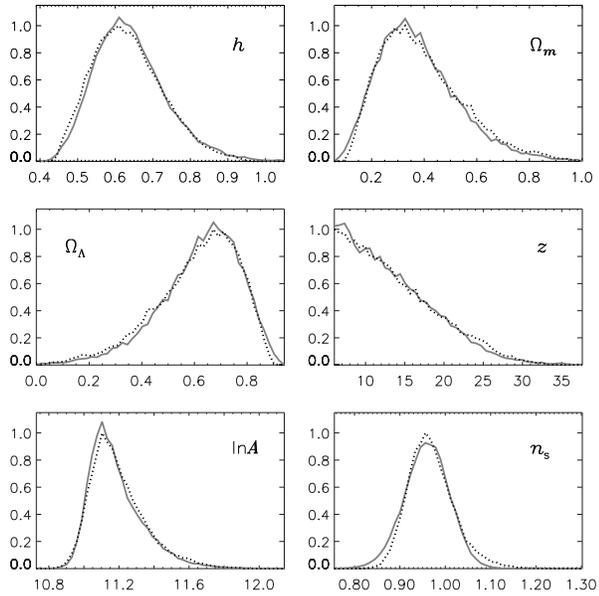}}}
\caption[]{\label{fig:1dcosmo}As in Fig.~\ref{fig:1dnorm} but
for other parameters.}
\end{figure}

Our normal fit to the likelihood can also be used to obtain
marginalized constraints on the cosmological parameters.
Marginalizations down to two dimensions and one dimension are shown in
Figures \ref{fig:2dcosmo} and \ref{fig:1dcosmo} respectively.  Note
that the highly non--Gaussian distributions of the cosmological
parameters are described well by the fit that is Gaussian in the
normal parameters.

\clearpage

\clearpage

The eigenmodes and eigenvalues $\lambda$ are shown in
Table~\ref{tab:eigenvectors}.  These are the eigenmodes of the {\em
fractional} Fisher matrix, $F^f_{ij} = \bar \theta_i F^a_{ij} \bar
\theta_j$ where $\bar \theta_i$ are the best--fit values of the normal
parameters.  We look at the fractional Fisher matrix so its
eigenvalues quantify the fractional error and not the absolute error.
The eigenvectors are ordered from highest eigenvalue to lowest
eigenvalue.  We see that current CMB data constrain four parameters at
the 10\% level or better.  As expected, the best--constrained
eigenvector is predominantly $\Theta_s$ and the worst--constrained is
almost entirely $z$.  The second--from--worst is $\omega_d$, though
even it is farily well--constrained with $\sqrt{\lambda}=0.16$.
Eigenvectors 2, 3 and 4 are mostly $t$, $A^*$ and $\omega_b$
respectively, though each with significant amounts of other parameters
mixed in.

We have also explored the likelihood distribution of the parameter
set proposed by \citet{kosowsky02}.  In 2-D projections, most pairs
of parameters do look Gaussian.  An exception is any pair involving
their variable ${\cal Z} \equiv \exp{(-2\tau)}$, which is
significantly less Gaussian than the variable we have used
to parameterize reionization:  the redshift of reionization, $z$.

\section{Discussion and Conclusions}

We have defined a set of parameters that, given current data,
have a likelihood that is normally distributed.  Of course,
our simple fit to the likelihood surface will very soon be
out-of-date due to the expected MAP data.  Still, we expect
the technique to continue to be useful.  Generally, as data
become more constraining, the Gaussian approximation becomes
better.

The challenge in the future will be to provide good parameterizations
for those parameters that remain poorly determined.
Some of these only have their effects at low $\ell$, such as
the parameters governing reionization.  For these it may be
useful to define parameters for which $\ln{C_l}$ is linear,
since the errors in this quantity are very nearly normally
distributed at low $\ell$ \cite{bond00}.  Other physical effects, such as
gravitational lensing, are only important at high $\ell$ where,
if instrument noise is the dominant contribution to error in $C_l$,
then $C_l$ will be nearly normally distributed.  For these one would
want to use parameterizations for which $C_l$ is linear.

An analytic form for the likelihood makes the actual application
of CMB parameter constraints to cosmological questions much easier.
One application is to quickly calculate constraints on various combinations
of the cosmological parameters such as $\sigma_8$ \citep{holder02}
or the age, $t_0$ \citep{knox01}.  Another is to combine the
CMB results with those from other probes to improve constraints
or search for inconsistencies \citep{wang01}.
Or one can forecast expected constraints from a combination of the CMB
and other probes such as supernovae \citep{frieman02} or galaxy
cluster number counts \citep{levine02}.  Although these applications
are possible without an analytic form for the CMB likelihood, such a
form greatly simplifies the analysis.

Use of normal parameters will also improve the efficiency of Monte Carlo
evaluations of the likelihood, by providing an analytic generating function
that is well--matched to the actual likelihood. Although the statistical
properties of the post--convergence chain
do not depend on the generating
function used, the time required for convergence depends
critically on the generating function. In particular the chain rapidly
converges if the generating function approximates the likelihood
well. Clearly, this issue becomes more important for larger parameter
spaces. A practical way to implement the use of normal parameters in
generating Monte Carlo Markov chains would be to create a small chain,
derive the normal parameters and the gaussian approximation to the
likelihood, and then use this gaussian likelihood as the generating
function.

In this paper we have explicitly given an analytic form for the
likelihood (given a certain data set) of cosmological parameters and
shown it to be an excellent approximation to the MCMC--calculated
likelihood. We view this as the final step in a long data-reduction
chain that begins with the time--ordered data.

\acknowledgments We thank N. Christensen, A. Kosowsky and R. Meyer
for useful conversations.  This work was supported in part by NASA grant
NAG5-11098.

\clearpage
\begin{table}
\begin{center}
\begin{tabular}{c|cccccc}
&         $\omega_b$ & $\omega_d$ & $z$    & $\Theta_s$ & $A^*$  & $t$  \\ \hline
$\omega_b$ &   1.000 &  -0.2270 &  -0.6840 & 0.3792   & 0.2572   & 0.05676 \\
$\omega_d$ & -0.2270 &    1.000 &  -0.03926& 0.008238 & -0.1875  & 0.2819  \\
$z$        & -0.6840 & -0.03926 &   1.000  & -0.2831  & 0.2031   & -0.1503 \\
$\Theta_s$ &  0.3792 & 0.008238 & -0.2831  &  1.000   & 0.003574 & 0.2755  \\
$A^*$      &  0.2572 &  -0.1875 & 0.2031   & 0.003574 & 1.000    & -0.8229 \\
$t$        & 0.05675 &   0.2819 & -0.1503  & -0.8229  & 0.2755   & 1.000   \\
\hline \\
$\sqrt{C_{ii}}$ & $0.0022$ &   0.023 &     0.26  & 0.0073 & 11.5 &  0.088 \\
mean            & 0.021 &     0.125 &      6.719 & 0.587 & 3.75  &  0.69
\end{tabular}
\end{center}
\caption{Correlation matrix, rms and mean of the normal fit to the
likelihood.}
\label{tab:normfit}
\end{table}

\begin{table}
\begin{center}
\begin{tabular}{c|cccccc|c}
eigenvector & $\omega_b$ & $\omega_d$ & $z$ & $\Theta_s$ & $A^*$ &  $t$ & $\lambda^{-1/2}$   \\ \hline
 1 &    0.02327 &   0.01554 &  -0.004625  &    0.9852  &   -0.1152  &    0.1240  &  0.0090\\
 2 &      0.3664 &    0.08785 &   -0.01180 &    -0.1448 &    -0.1890 &     0.8951& 0.015 \\
 3 &    -0.2624 &    -0.1551 &    0.03314 &    0.07245 &     0.8923 &     0.3232 & 0.048\\
 4 &      0.8759 &     0.1070 &   0.007523 &    0.05664 &     0.3731 &    -0.2811& 0.10 \\
 5 &     -0.1706 &     0.9779 &    0.01899 &   0.002654 &     0.1190 & -0.0003358 & 0.16\\
 6 &    0.009783 &   -0.01314 &     0.9992 & $-2.9\times 10^{-5}$ &   -0.03743 &   0.002542  & 2.8
\end{tabular}
\end{center}
\caption{Normal parameter Eigenvectors and the errors on their amplitudes, $\lambda^{-1/2}$.}
\label{tab:eigenvectors}
\end{table}
\clearpage

\bibliography{cmb3}

\begin{thebibliography}{29}
\expandafter\ifx\csname natexlab\endcsname\relax\def\natexlab#1{#1}\fi

\bibitem[{{Bartlett} {et~al.}(2000){Bartlett}, {Douspis}, {Blanchard}, \& {Le
  Dour}}]{bartlett00}
{Bartlett}, J.~G., {Douspis}, M., {Blanchard}, A., \& {Le Dour}, M. 2000,
  \aaps, 146, 507

\bibitem[{{Becker} {et~al.}(2001){Becker}, {Fan}, {White}, {Strauss},
  {Narayanan}, {Lupton}, {Gunn}, {Annis}, {Bahcall}, {Brinkmann}, {Connolly},
  {Csabai}, {Czarapata}, {Doi}, {Heckman}, {Hennessy}, {Ivezi{\' c}}, {Knapp},
  {Lamb}, {McKay}, {Munn}, {Nash}, {Nichol}, {Pier}, {Richards}, {Schneider},
  {Stoughton}, {Szalay}, {Thakar}, \& {York}}]{becker01}
{Becker}, R.~H., {Fan}, X., {White}, R.~L., {Strauss}, M.~A., {Narayanan},
  V.~K., {Lupton}, R.~H., {Gunn}, J.~E., {Annis}, J., {Bahcall}, N.~A.,
  {Brinkmann}, J., {Connolly}, A.~J., {Csabai}, I.~., {Czarapata}, P.~C.,
  {Doi}, M., {Heckman}, T.~M., {Hennessy}, G.~S., {Ivezi{\' c}}, {\v Z}.,
  {Knapp}, G.~R., {Lamb}, D.~Q., {McKay}, T.~A., {Munn}, J.~A., {Nash}, T.,
  {Nichol}, R., {Pier}, J.~R., {Richards}, G.~T., {Schneider}, D.~P.,
  {Stoughton}, C., {Szalay}, A.~S., {Thakar}, A.~R., \& {York}, D.~G. 2001,
  \aj, 122, 2850

\bibitem[{{Bennett} {et~al.}(1996){Bennett}, {Banday}, {Gorski}, {Hinshaw},
  {Jackson}, {Keegstra}, {Kogut}, {Smoot}, {Wilkinson}, \&
  {Wright}}]{bennett96}
{Bennett}, C.~L., {Banday}, A.~J., {Gorski}, K.~M., {Hinshaw}, G., {Jackson},
  P., {Keegstra}, P., {Kogut}, A., {Smoot}, G.~F., {Wilkinson}, D.~T., \&
  {Wright}, E.~L. 1996, \apjl, 464, L1

\bibitem[{{Benoit}(2002)}]{benoit02}
{Benoit}, A. e.~a. 2002, astro-ph/0210306

\bibitem[{{Best} {et~al.}(1995){Best}, {Cowles}, \& {Vines}}]{best95}
{Best}, N.~G., {Cowles}, M.~K., \& {Vines}, S.~K. 1995, {CODA: Convergence,
  Diagnosis, and Output Software for Gibbs Sampler Output} (version 0.30;
  Cambridge: MRC Biostatistics Unit)

\bibitem[{{Bond} {et~al.}(1998){Bond}, {Jaffe}, \& {Knox}}]{bond98}
{Bond}, J.~R., {Jaffe}, A.~H., \& {Knox}, L. 1998, \prd, 57, 2117

\bibitem[{{Bond} {et~al.}(2000){Bond}, {Jaffe}, \& {Knox}}]{bond00}
---. 2000, \apj, 533, 19

\bibitem[{{Christensen} {et~al.}(2001){Christensen}, {Meyer}, {Knox}, \&
  {Luey}}]{christensen01}
{Christensen}, N., {Meyer}, R., {Knox}, L., \& {Luey}, B. 2001, Classical
  Quantum Gravity, 18, 2677

\bibitem[{{Efstathiou} \& {Bond}(1999)}]{efstathiou99}
{Efstathiou}, G. \& {Bond}, J.~R. 1999, \mnras, 304, 75

\bibitem[{{Fan} {et~al.}(2002){Fan}, {Narayanan}, {Strauss}, {White}, {Becker},
  {Pentericci}, \& {Rix}}]{fan02}
{Fan}, X., {Narayanan}, V.~K., {Strauss}, M.~A., {White}, R.~L., {Becker},
  R.~H., {Pentericci}, L., \& {Rix}, H. 2002, \aj, 123, 1247

\bibitem[{{Frieman} {et~al.}(2002){Frieman}, {Huterer}, {Linder}, \&
  {Turner}}]{frieman02}
{Frieman}, J.~A., {Huterer}, D., {Linder}, E.~V., \& {Turner}, M.~S. 2002,
  astro-ph/0208100

\bibitem[{{Halverson} {et~al.}(2002){Halverson}, {Leitch}, {Pryke}, {Kovac},
  {Carlstrom}, {Holzapfel}, {Dragovan}, {Cartwright}, {Mason}, {Padin},
  {Pearson}, {Readhead}, \& {Shepherd}}]{halverson02}
{Halverson}, N.~W., {Leitch}, E.~M., {Pryke}, C., {Kovac}, J., {Carlstrom},
  J.~E., {Holzapfel}, W.~L., {Dragovan}, M., {Cartwright}, J.~K., {Mason},
  B.~S., {Padin}, S., {Pearson}, T.~J., {Readhead}, A.~C.~S., \& {Shepherd},
  M.~C. 2002, \apj, 568, 38

\bibitem[{{Hivon} {et~al.}(2002){Hivon}, {G{\' o}rski}, {Netterfield}, {Crill},
  {Prunet}, \& {Hansen}}]{hivon02}
{Hivon}, E., {G{\' o}rski}, K.~M., {Netterfield}, C.~B., {Crill}, B.~P.,
  {Prunet}, S., \& {Hansen}, F. 2002, \apj, 567, 2

\bibitem[{{Holder}(2002)}]{holder02}
{Holder}, G. 2002, astro-ph/0207600

\bibitem[{{Hu} {et~al.}(2001){Hu}, {Fukugita}, {Zaldarriaga}, \&
  {Tegmark}}]{hu01}
{Hu}, W., {Fukugita}, M., {Zaldarriaga}, M., \& {Tegmark}, M. 2001, \apj, 549,
  669

\bibitem[{{Hu} \& {White}(1997)}]{hu97}
{Hu}, W. \& {White}, M. 1997, \apj, 479, 568

\bibitem[{{Kaplinghat} {et~al.}(2002){Kaplinghat}, {Knox}, \&
  {Skordis}}]{kaplinghat02}
{Kaplinghat}, M., {Knox}, L., \& {Skordis}, C. 2002, \apj, 578, 665,
  astro-ph/0203413

\bibitem[{Kaplinghat {et~al.}(2002)}]{kaplinghat02a}
Kaplinghat, M. {et~al.} 2002, astro-ph/0207591

\bibitem[{{Knox} {et~al.}(2001){Knox}, {Christensen}, \& {Skordis}}]{knox01}
{Knox}, L., {Christensen}, N., \& {Skordis}, C. 2001, \apjl, 563, L95

\bibitem[{{Kosowsky} {et~al.}(2002){Kosowsky}, {Milosavljevic}, \&
  {Jimenez}}]{kosowsky02}
{Kosowsky}, A., {Milosavljevic}, M., \& {Jimenez}, R. 2002, \prd, 66, 63007

\bibitem[{{Lee} {et~al.}(2001){Lee}, {Ade}, {Balbi}, {Bock}, {Borrill},
  {Boscaleri}, {de Bernardis}, {Ferreira}, {Hanany}, {Hristov}, {Jaffe},
  {Mauskopf}, {Netterfield}, {Pascale}, {Rabii}, {Richards}, {Smoot},
  {Stompor}, {Winant}, \& {Wu}}]{lee01}
{Lee}, A.~T., {Ade}, P., {Balbi}, A., {Bock}, J., {Borrill}, J., {Boscaleri},
  A., {de Bernardis}, P., {Ferreira}, P.~G., {Hanany}, S., {Hristov}, V.~V.,
  {Jaffe}, A.~H., {Mauskopf}, P.~D., {Netterfield}, C.~B., {Pascale}, E.,
  {Rabii}, B., {Richards}, P.~L., {Smoot}, G.~F., {Stompor}, R., {Winant},
  C.~D., \& {Wu}, J.~H.~P. 2001, \apjl, 561, L1

\bibitem[{{Levine} {et~al.}(2002){Levine}, {Schulz}, \& {White}}]{levine02}
{Levine}, E.~S., {Schulz}, A.~E., \& {White}, M. 2002, \apj, 577, 569

\bibitem[{{Lineweaver} {et~al.}(1997){Lineweaver}, {Barbosa}, {Blanchard}, \&
  {Bartlett}}]{lineweaver97}
{Lineweaver}, C.~H., {Barbosa}, D., {Blanchard}, A., \& {Bartlett}, J.~G. 1997,
  \aap, 322, 365

\bibitem[{{Tegmark}(1997{\natexlab{a}})}]{tegmark97a}
{Tegmark}, M. 1997{\natexlab{a}}, \apjl, 480, L87+

\bibitem[{{Tegmark}(1997{\natexlab{b}})}]{tegmark97b}
---. 1997{\natexlab{b}}, \prd, 55, 5895

\bibitem[{{Tegmark} \& {Zaldarriaga}(2000)}]{tegmark00}
{Tegmark}, M. \& {Zaldarriaga}, M. 2000, \apj, 544, 30

\bibitem[{{Wandelt} {et~al.}(2001){Wandelt}, {Hivon}, \& {G{\'
  o}rski}}]{wandelt01}
{Wandelt}, B.~D., {Hivon}, E., \& {G{\' o}rski}, K.~M. 2001, \prd, 64, 83003

\bibitem[{{Wang} {et~al.}(2002){Wang}, {Tegmark}, \& {Zaldarriaga}}]{wang01}
{Wang}, X., {Tegmark}, M., \& {Zaldarriaga}, M. 2002, \prd, 65, 123001

\bibitem[{{Wright} {et~al.}(1996){Wright}, {Hinshaw}, \& {Bennett}}]{wright96}
{Wright}, E.~L., {Hinshaw}, G., \& {Bennett}, C.~L. 1996, \apjl, 458, L53+

\end{thebibliography}

\end{document}